\documentclass[twocolumn,prl,superscriptaddress,amsmath,amssymb]{revtex4} 
\usepackage[latin9]{inputenc}
\usepackage{amsmath}
\usepackage{graphicx}
\usepackage{esint}

\makeatletter
\@ifundefined{textcolor}{}
{%
 \definecolor{BLACK}{gray}{0}
 \definecolor{WHITE}{gray}{1}
 \definecolor{RED}{rgb}{1,0,0}
 \definecolor{GREEN}{rgb}{0,1,0}
 \definecolor{BLUE}{rgb}{0,0,1}
 \definecolor{CYAN}{cmyk}{1,0,0,0}
 \definecolor{MAGENTA}{cmyk}{0,1,0,0}
 \definecolor{YELLOW}{cmyk}{0,0,1,0}
}

\usepackage{color}
\usepackage{dcolumn}
\usepackage{bm}

\makeatother

\begin{document}

\title{Tunable topological phases with fermionic atoms in a one-dimensional flux lattice}

\author{Y. Deng}
\affiliation{State Key Laboratory of Low Dimensional Quantum Physics,Department of Physics, Tsinghua University, Beijing 100084, China}

\author{R. L\"u}
\author{L. You}
\email{lyou@mail.tsinghua.edu.cn}
\affiliation{State Key Laboratory of Low Dimensional Quantum Physics,Department of Physics, Tsinghua University, Beijing 100084, China}
\affiliation{Collaborative Innovation Center of Quantum Matter, Beijing 100084, China}

\date{\today}
\begin{abstract}
We present a simple scheme for implementing a one-dimensional (1D) magnetic-flux lattice of ultracold fermionic spin-$1/2$ atoms. The resulting tight-binding model supports gapped and gapless topological phases, and chiral currents for Meissner and vortex phases. Its single-particle spectra exhibit topological flat bands at small flux, and the flatness sensitively depends on hopping strength. An effective $p$-wave interaction arises in a $s$-wave paired superfluid. Treating atomic internal states as forming a synthetic dimension and balancing the interplay of magnetic flux and Zeeman field, our model describes
a tunable topological Fermi superfluid, which paves the way towards experimental explorations of non-Abelian topological matter in 1D atomic quantum gases.
\end{abstract}

\maketitle

\section{Introduction}
Topological superfluids are topical areas of research in quantum many-body physics. Among the diverse topics studied, the $p$-wave superfluid represents a paradigm, which hosts Majorana fermionic excitations with non-Abelian statistics~\cite{Ivanov01,Nayak08,Lutchyn10} essential to topological quantum computation~\cite{Kitaev03}. It remains to be realized experimentally, although, albeit the recent evidence for $p$-wave interaction in a nanowire topological superconductor~\cite{Mourik12,MTDeng,Das12}. Unlike solid-state systems (of electrons), interactions and the environment of
ultracold atoms are tunable or controllable~\cite{Bloch2008,Chin2010}, opening unparalleled opportunities for quantum simulations of topological superfluids. The recent breakthroughs of Raman-assisted tunneling for ultracold atoms in optical lattices ~\cite{Aidelsburger11,Kennedy15} establish concrete examples with strong synthetic magnetic fields
capable of exploring exotic states~\cite{Dalibard2011,Goldman2014}, and the successful realizations of spin-orbit coupling (SOC) interactions~\cite{Wu15,Huang15} support explorations of topological superfluids of ultracold atoms with $s$-wave interactions~\cite{Zhang08,Sato09,Zhai11}.

\begin{figure}[tbp]
\includegraphics[width=0.98\columnwidth]{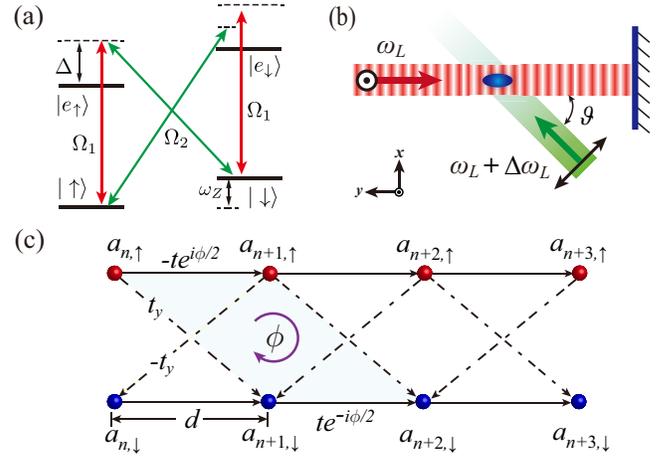}
\caption{(Color online). (a) The proposed atomic level diagram and Raman couplings, with a plane wave laser intersecting a $1$D standing-wave one at angle $\vartheta$ as in (b). (c) The resulting $1$D model describes synthetic magnetic-flux $\phi$ per plaquette (light blue region) with spatially dependent complex hopping along the physical dimension and staggered Raman-assisted spin-flip hopping along the synthetic dimension. }
\label{scheme}
\end{figure}

The difficulties with atomic gases lie at the increasingly complicated atom-atom as well as atom-light interactions required for realizing model systems of topological superfluids. In this paper, we present a simple implementation for a tunable flux lattice supporting topological states. It is based on Raman-assisted staggered spin-flip hopping for atoms in a one-dimensional (1D) optical lattice. Depending on the interplay of magnetic flux and spin-flip hopping, our model supports gapped and gapless topological phases and exhibits a quantum phase transition between the chiral vortex phase (VP) and the Meissner phase (MP).
Additionally, for atoms with $s$-wave interactions, a topological superfluid with zero-energy Majorana modes is predicted to exist. Compared with models requiring high-dimensional SOC ($>$ 1D) or high partial-wave ($l>0$) interactions, our system is built from simple laser configurations and with $s$-wave interactions and thus presents itself as a natural playground for exploring rich varieties of nontrivial topology and strongly correlated phenomena, e.g. the Su-Schrieffer-Heeger model~\cite{Atala13}, chiral  current~\cite{Atala14,Wei14,Hugel2014,Greschner15}, synthetic dimensions~\cite{Mancini15,Stuhl15,Celi14,Gra15}, fractional helical states~\cite{Zeng15,Barbarino15}, and topological charge pumping~\cite{Nakajima16,Lohse16}.

This paper is organized as follows. In Sec.~\ref{effham}, we introduce our model of Raman-assisted SOC and derive the single-particle Hamiltonian. In Sec.~\ref{edge}, we study the band topology and demonstrate the appearance of chiral currents for a synthetic 1D flux lattice in Sec.~\ref{current}. In Sec.~\ref{tisup}, we present the phase diagram and topological $p$-wave superfluids of the system. Finally, we give a brief summary in Sec.~\ref{cons}.

\section{Model and Hamiltonian}\label{effham}
$N$ fermionic atoms of level structure as illustrated in Fig.~\ref{scheme}(a) are subjected to a bias magnetic field ${\bf B}$ along the $z$-axis, e.g. for alkali $^{40}$K or $^{6}$Li atoms with two electronic excited states $|e_{\sigma}\rangle$ of $^2P_{1/2}$ and two ground hyperfine spin states $|\sigma\rangle$ of $^2S_{1/2}$ for $\sigma=\uparrow,\downarrow$. The specific Zeeman states that match the Raman selection rules are shown in Fig.~\ref{scheme}(b) with $\pi$-polarized transitions driven by a standing wave laser $|\sigma\rangle\leftrightarrow |e_{\sigma}\rangle$ of Rabi frequency $\Omega_1(y)=\Omega_1\cos(k_Ly)$ and $\sigma$-polarized transitions driven by a plane-wave laser $|\sigma\rangle\leftrightarrow |e_{\sigma'}\rangle$ ($\sigma \neq\sigma'$) with Rabi frequency $\Omega_2(y) = \Omega_{2}e^{-i\kappa y}$ and $\kappa=k_L \cos{\vartheta}$. $k_L$ is the laser wave vector, and ${\vartheta}$ is a tunable angle with respect to the $y$ axis.

If the differential detuning $\Delta \omega_L$ between the Raman lasers is near resonant with the Zeeman shift $\hbar \omega_Z$, states $|\!\uparrow\rangle$ and $|\!\downarrow\rangle$ from a closed subsystem well separated from the others when the quadratic Zeeman shift is sufficiently large~\cite{Huang15}. In the large detuning limit $|\Omega_{1,2}/\Delta|\ll1$,
the excited states $|e_{\sigma}\rangle$ can be eliminated adiabatically to give a 1D optical lattice ${\cal U}_{\rm{ol}}(y)= U_1\cos^2(k_Ly)$ with $U_{1,2} = \Omega^{1,2}_0/\Delta$ as the ac Stark shift and $d = \pi/k_L$ as the lattice constant of a unit cell. Neglecting terms with the high-frequency prefactor $e^{\pm i2\Delta\omega_{L}t}$ and rotating the spin basis by $\left|\uparrow\right\rangle \rightarrow e^{i\kappa y/2}\left|\uparrow\right\rangle$ and $\left|\downarrow\right\rangle \rightarrow e^{-i\kappa y/2}\left|\downarrow\right\rangle$, the single-particle Hamiltonian becomes
\begin{eqnarray}
{\boldsymbol{h}}=\frac{({\mathbf{p}}-{\mathbf{A}})^{2}}{2M}+\Omega\cos(k_Ly) \hat{\sigma}_{x}-\frac{\delta}{2}\hat{\sigma}_{z}+{\cal U}_{\rm{ol}}(y)\hat{I},\label{single-SO}
\end{eqnarray}
with $M$ as the atomic mass and $\hat{I}$ as the identity matrix. $\Omega=\Omega_{1}\Omega_{2}/\Delta$ is the effective Raman Rabi coupling, ${\mathbf{A}}=-\hbar\kappa\hat{\sigma}_{z}/2$ is an effective vector potential, $\delta=\omega_Z+\Delta \omega_L$ is the effective Zeeman field (two-photon detuning), and $\hat{\sigma}_{x,y,z}$'s are the Pauli matrices. Light-induced heating is reduced because an external optical lattice is not required~\cite{Deng2016}.

When $|\Omega_2/\Omega_1|\ll 1$ and the blue ($\Delta>0$) lattice potential is sufficiently strong, the above single atom interaction can be cast into a tight-binding model Hamiltonian with nearest-neighbor hopping
\begin{eqnarray}
{H}_0 &= &- t \sum_n
\left(\hat{a}^\dag_{n,\uparrow}  \hat{a}_{n+1,\uparrow} e^{-i\phi/2} +\hat{a}^\dag_{n,\downarrow}\hat{a}_{n+1,\downarrow}e^{i\phi/2}
+{\rm{H.c.}}\right) \nonumber \\
&&+ t_y \sum_n (-1)^n
\left(\hat{a}^\dag_{n,\uparrow}\hat{a}_{n+1,\downarrow}-
\hat{a}^\dag_{n,\uparrow}\hat{a}_{n-1,\downarrow} +
{\rm{H.c.}}\right)\nonumber \\
&& - \frac{\delta}{2} \sum_n
\left(\hat{a}^\dag_{n,\uparrow}\hat{a}_{n,\uparrow}
-\hat{a}^\dag_{n,\downarrow}\hat{a}_{n,\downarrow}\right),
\label{hamlat}
\end{eqnarray}
where $\hat{a}_{n,\sigma}$ is the atomic annihilation operator for the $n$th site and and $\phi=\kappa d$ is an easily tuned Peierls phase (magnetic flux) by changing ${\vartheta}$, $t=-\int d{{y}}\, w_{{{n}}}^{*}({{y}})\left[{\bf {p^{2}}}/(2M)+{\cal U}_{\text{ol}}({{y}})\right]w_{{{n+1}}}({{y}})$ is the nearest-neighbor spin-independent hopping with $w_{{n}}({y})$ as the Wannier function of the lowest $s$ orbit. The Raman-assisted nearest-neighbor spin-flip hopping $t_{y}\!=\!\Omega\int d{y}w_{n}^{*}({{y}})|\cos(k_Ly) |w_{n+1}({y})$ is staggered along the 1D lattice. Whereas the Raman-assisted on-site spin-flip hopping strength $\Omega\int d{y}w_{n}^{*}({{y}})\cos(k_Ly) w_{n}({y})$ is zero since the atoms are symmetrically localized at the nodes for the blue lattice potential. In contrast, a staggered Raman-assisted on-site spin-flip hopping emerges with the absence of the nearest-neighbor spin-flip hopping along the physical dimension for a red lattice potential.

The gauge transformation $\hat{a}_{{{n}},\downarrow}^{\dag}\rightarrow(-1)^{n+1} \hat{a}_{{{n+1}},\downarrow}^{\dag}$ eliminates the staggering factor~\cite{Liu14} and reduces Eq.~(\ref{hamlat}) to
\begin{eqnarray}
{H}_0 &= &- t \sum_n
\left(\hat{a}^\dag_{n,\uparrow}  \hat{a}_{n+1,\uparrow} e^{-i\phi/2}-\hat{a}^\dag_{n,\downarrow}\hat{a}_{n+1,\downarrow}e^{i\phi/2}
+{\rm{h.c.}}\right) \nonumber \\
& & + t_y \sum_n
\left(\hat{a}^\dag_{n,\uparrow}\hat{a}_{n+1,\downarrow}-
\hat{a}^\dag_{n,\uparrow}\hat{a}_{n-1,\downarrow} +
{\rm{h.c.}}\right)\nonumber \\
& & - \frac{\delta}{2} \sum_n
\left(\hat{a}^\dag_{n,\uparrow}\hat{a}_{n,\uparrow}
-\hat{a}^\dag_{n,\downarrow}\hat{a}_{n,\downarrow}\right),
\label{gau-hamlat}
\end{eqnarray}
whose corresponding schematic is depicted in Fig.~\ref{scheme}(c). Its $k$-space form easily is obtained as
\begin{eqnarray}
H_{0}({{k}})=\sum_{{{k}},\sigma\sigma'}\hat{a}_{{{k}} \sigma}^{\dag}\Big[\epsilon({{k}})\hat{I}+\sum_{\alpha=y,z}d_{\alpha}({{k}}) \hat{\sigma}_{\alpha}\Big]_{\sigma\sigma'}\hat{a}_{{{k}}\sigma'},\label{k-single}
\end{eqnarray}
where $\epsilon({{k}})=-2t\sin(\phi/2)\sin(kd)$, $d_{y}({{k}})\!=\!-2t_{y}\sin(kd)$, and $d_{z}({{k}})\!=\!-\delta/2-2t\cos(\phi/2)\cos(kd)$. Such a system of Eq.~(\ref{k-single}) belongs to the symmetry class $D$ which preserves particle-hole symmetry but breaks time-reversal symmetry. It supports a topological nontrivial ground state characterized by the 1D $Z_2$ invariant~\cite{AZ,Ludwig08}.

\begin{figure}[ht]
\includegraphics[width=0.98\columnwidth]{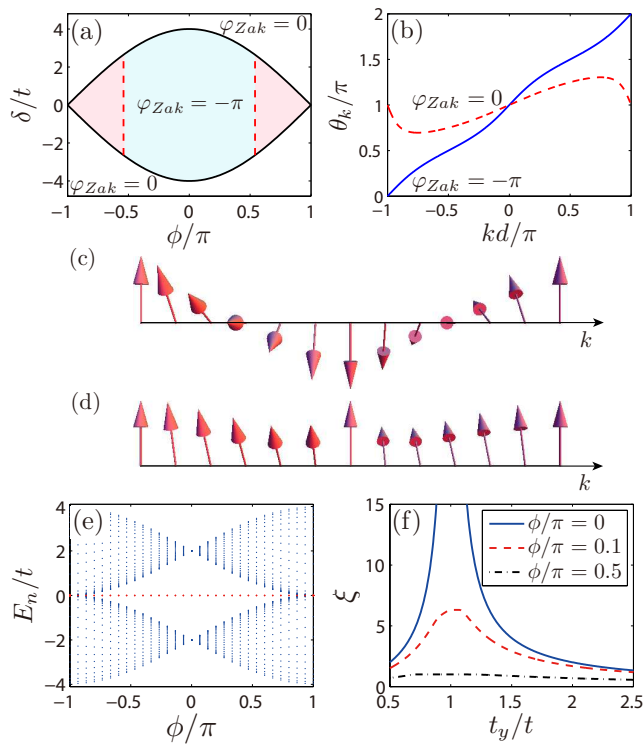}
\caption{(Color online). (a) The Zak phase $\varphi_{\rm{Zak}}$ on the $\phi$-$\delta$ parameter plane. The black-solid (red-dashed) line denotes the topological nontrivial-trivial (gapped-gapless) phase transition. (b) Typical results of phase $\theta_k$ for the topological phase at $\delta/t=0$ (blue-solid line) and the trivial phase at $\delta/t=4$ (red-dashed line).
(c) and (d) The Bloch vector $\hat{n}_{-,k}$ illustrated for the topological phase at $\delta/t=0$ and the trivial phase at $\delta/t=-4$. (e) The dependence of energy spectra $E_n$ on $\phi$ at $t_y/t=1$ and $\delta/t=0$. The red-dots denote the edge states. (f) The flatness parameter $\xi$ as a function of $t_y/t$ for different $\phi$'s with $\delta/t=0$.}
\label{sphase}
\end{figure}

\section{Band topology and edge states}\label{edge}
The single-particle spectra to the noninteracting Hamiltonian (\ref{k-single}) is given by $E_{\pm}({k}) =\epsilon({k}) \pm d(k)$ with $d(k)=\sqrt{d^2_y({k})+d^2_z({k})}$. The unit vectors $\hat{n}_{\pm,k} = \pm(0,\sin\theta_k,\cos\theta_k)$ for the upper ($+$) and lower ($-$) helicity branches are represented in the Bloch sphere
with $\theta_k=\arg[d_z(k)+id_y(k)]$. In the helicity basis, $H_{0}({{k}})$ is diagonalized and given by $H_{0}({{k}}) = \sum_{k}[E_{+}({k}) \hat{c}^\dag_{k+}\hat{c}_{k+}+E_{-}({k}) \hat{c}^\dag_{k-}\hat{c}_{k-}]$, with the corresponding fermionic atom annihilation operators,
\begin{eqnarray}
\hat{c}_{k+}&=& \cos(\theta_k/2) \hat{a}_{k\uparrow} -
i\sin(\theta_k/2) \hat{a}_{k\downarrow}, \nonumber \\
\hat{c}_{k-}&=& -i\sin(\theta_k/2) \hat{a}_{k\uparrow} +
\cos(\theta_k/2) \hat{a}_{k\downarrow}.
\label{helicity}
\end{eqnarray}
Their associated band topology is determined by the Zak phase for the lower branch $\varphi_{\rm{Zak}} =-\frac{1}{2}\int_{-G/2}^{G/2} \partial_k \theta_k dk$ with $G=2\pi/d$. Due to the $Z_2$ invariance of the system, the gauge dependent Zak phase can only take two distinct values for our choice of the unit cell.

Figure~\ref{sphase}(a) shows the phase diagram with $\varphi_{\rm{Zak}}=-\pi(0)$ for the topological (trivial) phase. At the topological phase transition, the bulk gap $E_{g}^{(d)}=2\min[d({k})]$ must be closed, which gives rise to the $t_y$-independent phase boundary $\delta/t=\pm 4\cos(\phi/2)$. Although it remains unchanged with diminishing $t_y$, the value of $E_{g}^{(d)}$ depends on $t_y$. In the absence of the Zeeman field ($\delta=0$), the system stays topological with the bulk gap closing only occurring at $\phi/\pi=\pm1$. The flux-induced asymmetric potential $\epsilon({k})$, however, may induce a closing of the indirect bulk gap $E_{g}^{(i)}=2\min[E_+(k)]$ accompanied by an inverse shift of the extreme points for both the upper and the lower bands~\cite{Deng2016}. The system therefore can enter into a gapless topological phase at $E_{g}^{(i)}=0$ where the gapped-to-gapless phase transition satisfies $|\sin(\phi/2)|=t_{y}/t$. The gapless phase vanishes once $t_{y}\geq t$.

The phase $\theta_k$ and the Bloch vector $\hat{n}_{-,k}$ for different phases are plotted in Figs.~\ref{sphase}(b)-(d) for $t_y/t=1$ and $\phi=\pi/2$. We can visualize $\hat{n}_{-,k}$ for a typical topological phase characterized by the spin texture
with $2\pi$ phase winding~\cite{Hugel2014}, which exhibits the topologically protected twofold degenerate edge modes by imposing a hard-wall confinement along the lattice direction [Fig.~\ref{sphase}(e)]. The bulk gap is shown clearly to be monotonically decreasing with increasing flux $|\phi|$. Meanwhile,  the bandwidth of the lower branch, $E_{\rm{bw}}=\max[E_-(k)]-\min[E_-(k)]$, with respect to $t_y$ is dominated by $\phi$. The system exhibits a large flatness ratio $\xi=E_{g}^{(d)}/E_{\rm{bw}}$ which sensitively depends on $t_y/t$ indicated by the sharp peak structure at small $\phi$ as shown in Fig~\ref{sphase}(d). The measured band structure thus potentially allows for a precision determination of the hopping strength. Different from the small $\phi$ limit, $\xi$ becomes insensitive to $t_y/t$ at large $\phi$.

\begin{figure}[ht]
\includegraphics[width=0.9\columnwidth]{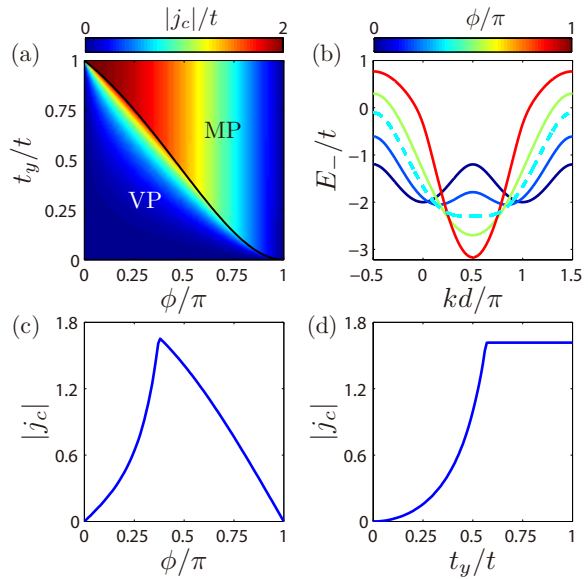}
\caption{(Color online). (a) The phase diagram for our model on the $\phi$-$t_y$ parameter plane at $\delta/t=0$. The color with the blue-red gradient shading indicates the values of $|{j}_c|$. (b) The typical lower branch $E_-$ for different $\phi$'s at $t_y/t=0.6$. The dashed line characterizes the VP and MP phase transitions. Chiral current $|{j}_c|$ as a function of (b) $\phi$ with $t_y/t=0.6$ and (c) $t_y/t$ with $\phi/\pi=0.4$, respectively.}
\label{cphase}
\end{figure}

\section{Chiral currents}\label{current}
The presence of magnetic flux as shown in Fig.~\ref{scheme}(c) results in a gauge-invariant net chiral current, which takes the form $j_c= \frac{1}{N_l}\sum_n\langle j_{n,\uparrow} -j_{n,\downarrow} + j_{n,\perp}\rangle$
and is composed of an intraspecies component along the legs $j_{n, \sigma} = {it\zeta} \hat{a}^\dag_{n,\uparrow} \hat{a}_{n+1,\uparrow} e^{i\zeta \phi/2}+ \rm{ H.c.}$ and an interspecies component along the rungs $j_{n,\perp}= {it_x}(\hat{a}^\dag_{n,\uparrow}  \hat{a}_{n+1,\downarrow} - \hat{a}^\dag_{n,\uparrow} \hat{a}_{n-1,\downarrow})+\rm{ H.c.}$, with $\zeta=-1$ (1) for spin-$|\uparrow\rangle$ ($|\downarrow\rangle$) and $N_l$ being the total number of lattice sites. The chiral current reduces to a Peierls phase of $\phi$ for the ground-state energy $E_g$ with $j_c=2\partial E_g/\partial \phi$~\cite{Greschner15}. It is measurable using spin-selective imaging of the lattice momentum distribution~\cite{Mancini15}.

Figure~\ref{cphase}(a) shows the current strength $|j_c|$ as a function of the hopping ratio $t_y/t$ and flux $\phi$. Tuning the lower helicity band structure~\cite{Atala14}, one induces a quantum phase transition between the VP and the MP where the former phase displays the same periodic modulation current whereas the latter shows a uniform current. This phase transition corresponds to the merging of two local minima of $E_-(k)$ into a single one as shown in Fig.~\ref{cphase}(b). At the phase boundary (black-solid line), the critical value of $t_y^{(c)}/t$ obeys $t^{(c)}_y/t=[\sqrt{1+3\cos^2(\phi/2)}-\sin(\phi/2)]/2$. It decreases gradually from $1$ to $0$ when the flux $\phi$ increases from $0$ to $\pi$. In contrast to the chiral current in two-leg bosonic ladders~~\cite{Atala14}, the VP favors small flux $\phi$ and hopping strength $t_y$. As shown in Fig.~\ref{cphase}(c), the current displays a peak structure which reaches maximum at the boundary of the VP-MP transition. Another typical feature of $|j_c|$ is that the system belongs to VP when $t_y<t_y^{(c)}$, with $j_c$ rapidly growing with increasing $t_y$. But for $t_y>t_y^{(c)}$, the system changes into the MP with the current saturating to $j_c=-2t\cos(\phi/2)$ [Fig.~\ref{cphase}(d)]. These characteristic behaviors of $j_c$ for different phases can be used to monitor the VP-to-MP transition.

\section{Topological $p$-wave superfluid}\label{tisup}
Topological superfluids arise for atoms with gauge-invariant $s$-wave interactions. By introducing the conventional order parameter $\Delta_{s}=(U_{0}/N_l)\sum_{{k}}\langle\hat{a}_{{-k},\downarrow}\hat{a}_{{ k},\uparrow}\rangle$~\cite{qzero}, the mean-field interaction Hamiltonian becomes ${H}_{\text{int}} = \Delta_{{s}} \sum_{k}\left(\hat{a}_{{k},\uparrow}^\dag \hat{a}_{{-k},\downarrow}^\dag +{\rm{H.c.}}\right) - \frac{N_l}{U_0}|\Delta_{s}|^2$, where $U_{0}$ is the attractive interaction strength. Projected into the helicity basis, ${H}_{\text{int}}$ becomes
\begin{eqnarray}
{{H}}_{\text{int}} &=& \sqrt{\frac{\pi}{3}}\Delta_s \sum_k\left[ iY_{10}(\frac{\pi}{2}-\theta_k)(\hat{c}^\dag_{k+}\hat{c}^\dag_{-k+}- \hat{c}^\dag_{k-}\hat{c}^\dag_{-k-}) \right. \nonumber \\ &&\left.+2Y_{10}(\theta_k)\hat{c}^\dag_{k+}\hat{c}^\dag_{-k-}+\rm{H.c.}\right]- \frac{N_l}{U_0}|\Delta_{s}|^2,\label{pwave}
\end{eqnarray}
where $Y_{10}(\theta_k)$ is a spherical harmonics of rank 2, whereas the first two terms correspond to pairings of atoms with the same or opposite helicities assuming effective $p$-wave interactions from the $s$ wave. These interactions drastically change the single-particle topology of Eq.~(\ref{k-single}) and give rise to zero-energy Majorana modes, whose origins are different from the $p$-wave interaction proposed in Ref.~\cite{Wang16} that requires a spin-dependent optical lattice.

In the Nambu space with the operator $\hat{\Psi}_{{{k}}}=(\hat{a}_{{ {k},\uparrow}},\hat{a}_{{{k},\downarrow}},\hat{a}_{{{-k},\uparrow}}^{\dagger}, \hat{a}_{{{-k},\downarrow}}^{\dagger})^{T}$, the Hamiltonian becomes ${H} = \frac{1}{2}\sum_{k} \hat{\Psi}_{{k}}^\dag{\mathcal H}_{\text{BdG}}\hat{\Psi}_{{k}} - \frac{N_l}{U_0}|\Delta_{s}|^2 +\sum_{{k}} \xi_k$, with $\xi_k=\epsilon(k) -\mu$ and $\mu$ as the chemical potential. The corresponding Bogliubov-de Gennes (BdG) Hamiltonian that preserves the inherent particle-hole symmetry becomes
\begin{eqnarray}
{\mathcal{H}}_{\text{BdG}}(k)
&&=\left(\begin{array}{cc}
{\mathcal K}(k) & i\Delta_{s}\hat{\sigma}_y \\
-i\Delta^*_{s}\hat{\sigma}_y & -{\mathcal K}^*({
-k})
\end{array}%
\right), \label{BdG}
\end{eqnarray}%
with ${\mathcal K}({k})= [\epsilon({{k}}) -\mu] \hat{I} + d_{y}(k)\hat{\sigma}_y+ d_{z}(k)\hat{\sigma}_z$. By diagonalizing the BdG Hamiltonian, one finds the eigenequation satisfying ${\mathcal{H}}_{\text{BdG}}({k}){\psi}_{\eta}^{\beta}(k)= E_{\eta}^{\beta}(k){\psi}_{\eta}^{\beta}(k)$, with ${\psi}^{\beta}_{\eta,{k}}=[u^{\beta}_{\eta,{k},\uparrow},u^{\beta}_{\eta,{ k},\downarrow},v^{\beta}_{\eta,{k},\uparrow},v^{\beta}_{\eta,{k},\downarrow}]^T$ as the wavefunctions and $E^{\beta}_{\eta,{k}}$ as the eigenenergies of the Bogoliubov quasiparticles. The index $\beta=+$ ($-$) represents the particle (hole) band and $\eta=1$ ($2$) denotes the upper (lower) helicity branch. The topology for the interacting system is characterized by a 1D $Z_2$ number for the hole branch~\cite{Takeshi13} $\nu = \frac{i}{\pi} \sum_{\eta=1,2} \int_{-G/2}^{G/2}\langle\psi_{\eta}^-(k)|\partial_k \psi_{\eta}^-(k)\rangle d k$, with $\nu=1(0)$ denoting topological (trivial) phases.

\begin{figure}[ht]
\includegraphics[width=0.95\columnwidth]{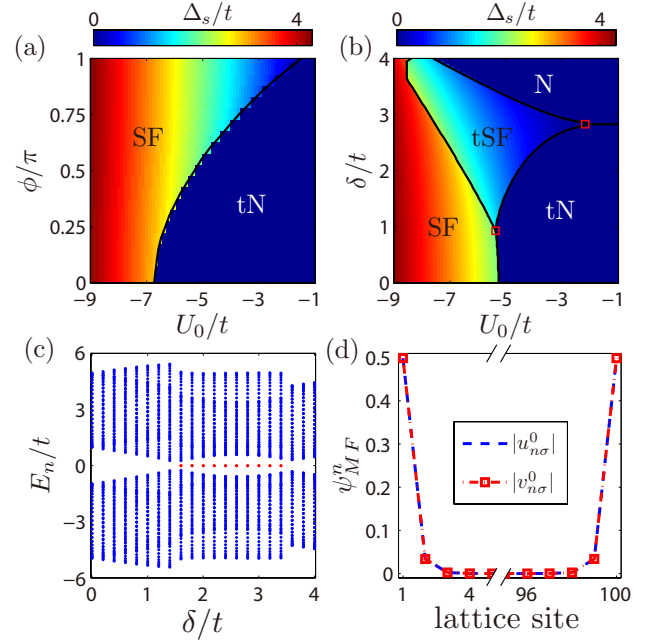}
\caption{(Color online). The phase diagram on the $U_0$-$\phi$ plane at $\delta/t=0$ (a) and on the $U_0$-$\delta$ plane at $\phi/\pi=0.5$ (b), with the blue-red color gradient shading indicating the value of $\Delta_s/t$. The red square points in (b) represent the two tricritical points of the quantum phase transitions. (c) The typical quasiparticle spectra $E_n$ as a function of $\delta$ with edge modes denoted by the red-dots. (d) The spatial distribution of the quasiparticle wave-functions for the Majorana modes along the lattice site at $\delta/t=2.5$. For (c) and (d), other parameters are $\phi/\pi=0.5$ and $U_0/t=-6$.}
\label{mphase}
\end{figure}

Figures~\ref{mphase}(a) and ~\ref{mphase}(b) summarize the corresponding phase diagrams, with the superfluid phase labeled by ``SF", the normal phase ($\Delta_s=0$) by ``N", and the topological phase for the particle ($\varphi_{\rm{Zak}}=-\pi$) or quasiparticle ($\nu=1$) by ``t". The phase boundaries are characterized by $\Delta_s$ and $\varphi_{\rm{Zak}}$ ($\nu$) with the self-consistent equation for $\Delta_s$ solved numerically using analytical derivatives by the Green's function method~\cite{Deng2016} $\Delta_{s} = \frac{U_0}{N_l}\sum_{k}\left[f(\mathcal{H}_{\text{BdG}}({ k}))\right]_{14}$, where $f(\cdot)$ denotes the Fermi-Dirac distribution. The 1D BdG approach is expected to capture the qualitative topological features when interactions are not too strong~\cite{Haller09}. Limited by the validity regime of our theory, we take $|U_0/t|\leq 9$ in the numerical simulation and with other parameters as $t_{y}/t=1$ and $\mu=1$ for half filling.

The topological phase transition is associated with the critical pairing order parameter $\Delta^{(c)}_{s}=|\delta/2\pm2t\cos(\phi/2)|$. As shown in Fig.~\ref{mphase} (a), for $\Delta^{(c)}_{s}>0$, the value of $\Delta_s$ does not smoothly decrease to zero, and a finite threshold $\Delta^{(\rm{th})}_s$ exists at the boundary of the tN-to-SF transition. As expected, $\Delta^{(\rm{th})}_s$ for the emergence of the SF as well as $\Delta^{(c)}_{s}$ monotonically decreases as $\phi$ increases, corresponding to the parameter region for the SF largely enhanced. However, the tSF state which possesses a nonzero 1D topological number is absent since $\Delta^{(\rm{th})}_s >\Delta^{(c)}_{s}$, which indicates that the tSF state favors a small threshold of superfluidity.

Further exploration of the tSF state leads to a more interesting quantum phase as depicted in Fig.~\ref{mphase}(b) with a significantly reduced $\Delta_s^{(\rm{th})}$ due to the interplay of $U_0$ and $\delta$. In particular, we note a tSF state with an intermediate value of $\Delta_s$. When the Zeeman field is over the first tricritical point ($\delta/t>0.93$), the parameter region for the tSF state gradually increases, until it reaches the second tricritical point of $\delta/t=2\sqrt{2}$,
after which it decreases approximately linearly with further increasing of $\delta$. Compared with the SF state, $\Delta_s$ for the tSF state more sensitively depends on $\delta$. In addition, there exists a tN-to-N transition at $\delta/t=2\sqrt{2}$,
which is independent of the interaction strength when $|U_0|<2.2$.

Figure~\ref{mphase}(c) shows the edge states with two-fold degenerate zero-energy modes for the tSF state as predicted due to the bulk-edge correspondence. The quasiparticle operator for each zero-energy mode is found to be expandable as $\hat{\gamma}_0=\sum_{n\sigma}u^0_{n\sigma}\hat{a}_{n\sigma} +v^0_{n\sigma}\hat{a}_{n\sigma}^\dag$, with the wavefunctions
well localized at the boundary [Fig.~\ref{mphase}(d)] and take symmetric forms $u^0_{n\sigma}= (v^0_{n\sigma})^*$. As a consequence, two localized Majorana modes are confirmed for the tSF phase, each corresponds to the antiparticle of itself or $\hat{\gamma}_0=\hat{\gamma}_0^\dag$. These emergent Majorana modes for $s$-wave interaction atoms are well understood by the model Hamiltonian (\ref{pwave}) for the effective $p$-wave superfluids.

\section{Conclusions}\label{cons}
A simple 1D lattice model with magnetic flux and $p$-wave interaction is proposed based on spin-$1/2$ atoms possessing $s$-wave interactions. This 1D model supports a variety of quantum phases including the VP, MP, and tSF states
possessing zero-energy Majorana modes. The VP-to-MP phase transition and the topological $p$-wave superfluids are tunable by the interplay of synthetic flux and Zeeman field. The band structure and topology can be measured by mature techniques, such as momentum-resolved rf spectroscopy~\cite{Stewart08} and Bloch band topology mapping~\cite{Duca15,Aidelsburger15}. With slight modifications, our model can be extended to study topological Fulde-Ferrell-Larkin-Ovchinnikov states with population imbalance for fermionic superfluidity~\cite{Liao2010}.

\section*{ACKNOWLEDGMENTS}
Y.D acknowledges support by the NSFC (Grant No. 11604178) and by China Postdoctoral Science Foundation. R.L. acknowledges support from the NSFC (Grant No. 11274195). L.Y was supported by the MOST 973 Program (Grant No. 2013CB922004) of the National Key Basic Research Program of China and by the NSFC (Grants No. 91121005 and No. 91421305).


\begin{thebibliography}{10}
\bibitem{qzero} We verified that only a standard BCS superfluid concentrated on the zero center-of-mass momentum emerges at half filling.
\end{thebibliography}
\end{document}